\documentclass[nofootinbib,a4paper,aps,prd,10pt,superscriptaddress,reprint,showkeys,showpacs]{revtex4-1}
\usepackage{graphicx,amsmath,amssymb}
\usepackage{times}
\usepackage{amsfonts}
\usepackage{amsmath}
\usepackage{amssymb}
\usepackage{natbib}
\usepackage{graphicx}
\usepackage{enumitem}
\usepackage{afterpage}
\usepackage{balance}

\usepackage[colorlinks]{hyperref}
\usepackage[outdir=./Images/]{epstopdf}
\usepackage[export]{adjustbox}

\usepackage{hyperref}
\hypersetup{%
    ,urlcolor=blue
    ,citecolor=blue
    ,linkcolor=blue
    }

\newcommand{\qsubrm}[2]{{#1}_{\scriptscriptstyle{\textrm{#2}}}}

\def\be{\begin{equation}}
\def\ee{\end{equation}}
\def\bea{\begin{eqnarray}}
\def\eea{\end{eqnarray}}
\def\bse{\begin{subequations}}
\def\ese{\end{subequations}}

\graphicspath{{./Images/}}

\def\aap{A\&A}
\def\aj{AJ}
\def\apjl{ApJL}
\def\apjs{ApJS}

\def\jcap{JCAP}
\def\mnras{MNRAS}
\def\physrep{Physics Reports}

\begin{document}
\title{Do cosmological data rule out \texorpdfstring{$f(\mathcal{R})$}{f(R)} with \texorpdfstring{$w\neq-1$?}{w}}

\author{Richard~A.~\surname{Battye}}%
\email[]{richard.battye@manchester.ac.uk}
\affiliation{%
Jodrell Bank Centre for Astrophysics, School of Physics and Astronomy, The University of Manchester, Manchester, 
M13 9PL, U.K.
}
\author{Boris~\surname{Bolliet}}
\email[]{boris.bolliet@manchester.ac.uk}
\affiliation{%
Jodrell Bank Centre for Astrophysics, School of Physics and Astronomy, The University of Manchester, Manchester, 
M13 9PL, U.K.
}
\author{Francesco~\surname{Pace}}
\email[]{francesco.pace@manchester.ac.uk}
\affiliation{%
Jodrell Bank Centre for Astrophysics, School of Physics and Astronomy, The University of Manchester, Manchester, 
M13 9PL, U.K.
}

\label{firstpage}

\date{\today}

\begin{abstract}
We review the Equation of State (EoS) approach to dark sector perturbations and apply it to $f(\mathcal{R})$ gravity 
models of dark energy. We show that the EoS approach is numerically stable and use it to set observational constraints 
on designer models. Within the EoS approach we build an analytical understanding of the dynamics of cosmological 
perturbations for the designer class of $f(\mathcal{R})$ gravity models, characterised by the parameter 
$\qsubrm{B}{0}$ and the background equation of state of dark energy $w$. When we use the Planck cosmic microwave 
background temperature anisotropy, polarisation and lensing data as well as the baryonic acoustic oscillation data from 
SDSS and WiggleZ, we find $\qsubrm{B}{0}<0.006$ (95\% C.L.) for the designer models with $w=-1$. Furthermore, we find 
$\qsubrm{B}{0}<0.0045$ and $|w+1|<0.002$ (95\% C.L.) for the designer models with $w\neq -1$. Previous analyses found 
similar results for designer and Hu-Sawicki $f(\mathcal{R})$ gravity models using the Effective Field Theory (EFT) 
approach 
[Raveri {\it et~al.}, \href{https://doi.org/10.1103/PhysRevD.90.043513}{Phys. Rev. D {\bf 90}, 043513 (2014)}; 
Hu {\it et~al.}, \href{https://doi.org/10.1093/mnras/stw775}{Mon. Not. R. Astron. Soc. {\bf 459}, 3880 (2016)}]; 
therefore this hints for the fact that generic $f(\mathcal{R})$ models with $w\neq-1$ can be tightly constrained by 
current cosmological data, complementary to solar system tests 
[Brax {\it et~al.}, \href{https://doi.org/10.1103/PhysRevD.78.104021}{Phys. Rev. D {\bf 78}, 104021 (2008)};
Faulkner {\it et~al.}, \href{https://doi.org/10.1103/PhysRevD.76.063505}{Phys. Rev. D {\bf 76}, 063505 (2007)}]. 
When compared to a $w$CDM fluid with the same sound speed, we find that the equation of state for $f(\mathcal{R})$ 
models is better constrained to be close to -1 by about an order of magnitude, due to the strong dependence of the 
perturbations on $w$.
\end{abstract}

\pacs{04.50.Kd, 95.36.+x, 98.80.-k}

\keywords{Cosmology; modified gravity; dark energy; f(R) gravity}

\maketitle

\section{Introduction}
With the observational campaign of Supernovae type Ia \citep{Riess1998,Perlmutter1999,Riess2004,Riess2007}, 
followed by observations of the Cosmic Microwave Background (CMB) anisotropy \citep{Planck2016_XIII,Hinshaw2013}, the 
Baryon Acoustic Oscillations (BAO) \citep{Percival2010,Parkinson2012} and large scale structure  
\citep{Alam2017,Rota2017,Ata2018}, it has become widely accepted that the expansion of the universe is accelerating. 
The current observational data is consistent with the standard $\Lambda$ cold dark matter (CDM) model, where the 
accelerated expansion is caused by the cosmological constant $\Lambda$, and indicates no statistically significant 
evidence for dark energy and modified gravity models (see, e.g., \cite{Planck2016_XIV} and references therein).

Nevertheless, the cosmological constant suffers from important conceptual issues when it is interpreted in the context 
of quantum field theory (see, e.g., \cite{Weinberg2013} for a recent review). This has led part of the community to 
question the physical origin of the accelerated expansion and to investigate dark energy and modified gravity models 
(see, e.g., \cite{Clifton2012}). Whether these models do not suffer the same type of issues as the cosmological 
constant often remains under debate.

Moreover, the forthcoming galaxy surveys and stage IV CMB experiments will measure the acceleration of the universe and 
its consequences on structure formation at a level of accuracy never achieved before. Hence, research on dark energy 
and modified gravity is well motivated by the following question: In the light of this forthcoming data, will the 
cosmological constant still be the best answer to cosmic acceleration? In other words, is there a modified gravity 
or dark energy model that will account for the observational data in a better way than the cosmological constant? Of 
course, this has to be formulated in a precise statistical manner, see \cite{Giannantonio2015} for an example in 
the context of inflationary models.

Recently, the Horndeski models \citep{Horndeski1974,Deffayet2011,Kobayashi2011} have received a growing attention due 
to their generality. They include a scalar field coupled to gravity. The Horndeski Lagrangian is the most general one 
that leads to second order equations of motion for the scalar field. It is fully represented by four arbitrary time 
dependent functions of the scalar field and its kinetic term. 
Notable subclasses of the Horndeski models, obtained by specifying the unknown functions, are Quintessence 
\citep{Ford1987,Peebles1988,Ratra1988a,Wetterich1988,Caldwell1998,Copeland1998,Steinhardt1999,Barreiro2000}, 
$k$-essence \citep{ArmendarizPicon1999,Chiba2000,Hamed2004,Piazza2004,Scherrer2004,Mukhanov2006}, Brans-Dicke theory 
\citep{Brans1961,DeFelice2010a}, Kinetic Gravity Braiding (KGB) \citep{Deffayet2010,Pujolas2011} and $f(\mathcal{R})$ 
models \citep{Song2007,Silvestri2009,Sotiriou2010,DeFelice2010}. 
The latter can also be constructed by replacing the Ricci scalar $\mathcal{R}$ in the Einstein-Hilbert Lagrangian by an 
arbitrary function, $f(\mathcal{R})$, and are the main focus of this paper.

Here, we are interested in $f(\mathcal{R})$ models that mimic the $\Lambda$CDM (or the $w$CDM) cosmological expansion 
history but differ at the level of the dynamics of cosmological perturbations. Different approaches have been 
developed to study the phenomenology of cosmological perturbations in dark energy and modified gravity in a unified 
way, with the ultimate objective of deriving observational constraints. These include the Parameterized Post 
Friedmaniann (PPF) approach \citep{Skordis2009a,Baker2011,Baker2013,Ferreira2014}, the Equation of State for 
perturbations (EoS) approach \citep{Battye2012,Battye2013,Battye2014} (see also \citep{Kunz2007} for an earlier and 
similar approach), the Effective Field Theory (EFT) approach 
\citep{Bloomfield2013a,Bloomfield2013b,Gleyzes2013,Gleyzes2014,Gleyzes2015} and \cite{Bertacca2012} for an alternative 
method. 
They are in principle equivalent (see \citep{Bellini2018} for a numerical consistency analysis), although they differ 
with respect to the choices of the phenomenological parametrisation of dark energy and modified gravity. 
So far, the EFT approach has been applied to generic Horndeski models \citep{Gleyzes2013,Gleyzes2014}, while the EoS 
approach has been applied specifically to quintessence, $k$-essence and KGB models \citep{Battye2014}, $f(\mathcal{R})$ 
gravity \citep{Battye2016a} and Generalised Einstein-Aether theories \citep{Battye2017}. In this paper we use the EoS 
approach, for which the dark energy and modified gravity models are specified in terms of the anisotropic stress and 
pressure of the perturbed dark energy fluid.

The paper is organised as follows. In section~\ref{sect:EoS} we review the EoS approach and its numerical 
implementation in a Boltzmann code for arbitrary dark energy and modified gravity models. In Sec.~\ref{sect:fR} we 
recall the features of the designer $f(\mathcal{R})$ models that are relevant to our analysis. 
In Sec.~\ref{sect:results} we study the phenomenology of cosmological perturbations propagating in the dark energy 
fluid of the models with constant $\qsubrm{w}{de}$, numerically as well as analytically. In Sec.~\ref{sect:obs} we 
present the linear matter power spectrum, the CMB temperature angular anisotropy power spectrum and the CMB power 
spectrum of the lensing potential, computed for several designer models and we derive observational constraints on the 
free parameters of the designer models, i.e., $\qsubrm{w}{de}$ and $\qsubrm{B}{0}$, from current CMB and BAO data. 
In Sec.~\ref{sect:wCDM} we compare $f(\mathcal{R})$ and $w$CDM gravity models and their observational 
constraints. We discuss our results and conclude in Sec.~\ref{sect:conclusion}. In Appendix we present a comparison 
between the perturbed equations of state obtained within the EoS \citep{Battye2016} and EFT 
approaches \citep{Gleyzes2013,Gleyzes2014}.

Unless otherwise stated, we use $8\pi\mathcal{G}=1$, throughout the paper.

\section{Numerical implementation of the equation of state approach}\label{sect:EoS}
In the EoS approach, modifications to general relativity are written in the right hand side of the field equations. 
Then, they can be interpreted as a stress energy tensor, mapping any modified gravity theory to a corresponding dark 
energy fluid. More precisely, we have
\be
 G_{\mu\nu}=T_{\mu\nu}+D_{\mu\nu}\;, \label{eq:fe}
\ee
where $G_{\mu\nu}$ is the Einstein tensor, $T_{\mu\nu}$ is the stress energy tensor of the matter components, i.e., 
baryonic matter, radiation and dark matter, and $D_{\mu\nu}$ is the stress-energy tensor of the dark energy fluid. The 
background geometry is assumed to be isotropic and spatially flat, with a line element 
$ds^2 = -dt^2 + a^2 \delta_{ij} dx^idx^j$, where $a$ is the scale factor and $t$ is the cosmic time. Due to the Bianchi 
identities and the local conservation of energy for the matter components, the stress energy tensor of the dark sector 
is covariantly conserved,
\be
 \nabla^\mu D_{\mu\nu}=0\;. \label{eq:bi}
\ee

The linear perturbation of the conservation equations \eqref{eq:bi} yields the general relativistic version of the 
Euler and continuity equations for the velocity and density perturbation. They characterise the dynamics of 
cosmological perturbations and can be written in terms of a gauge invariant density perturbation, $\Delta$, and a 
rescaled velocity perturbation, $\Theta$. These two quantities are defined as
\be
 \Delta \equiv \delta+3(1+w)H\theta\;, \quad \Theta \equiv 3(1+w)H\theta\;,
\ee
where $w\equiv P/\rho$ is the background equation of state, $\rho$ and $P$ are the homogeneous density and pressure, 
$\delta\rho$ is the density perturbation, $\theta$ is the divergence of the velocity perturbation, and 
$H\equiv (d\ln{a}/dt)$ is the Hubble parameter.

The rescaled velocity perturbation, $\Theta$, is not a gauge invariant quantity, in the sense that its value depends on 
the choice of the coordinate system, see, e.g. \cite{Ma1995}. To see this, say that $\Theta$ is evaluated in the 
conformal Newtonian gauge (CNG), i.e., $\Theta^{\rm c}=\Theta$, where the superscript $\mathrm{c}$ indicates the CNG. 
Then the value of the rescaled velocity perturbation in the synchronous gauge (SG), $\Theta^{\rm s}$, is given, in 
Fourier space, by
\be
 \Theta^{\rm s} = \Theta^{\rm c}-3(1+w)T\;,\label{eq:S-C}
\ee
with
\be
 T\equiv\begin{cases}
 (h^\prime+6\eta^\prime)/(2\mathrm{K}^2) & \mathrm{in}\,\,\mathrm{the}\,\,\mathrm{SG}\;, \\
 0 & \mathrm{in}\,\,\mathrm{the}\,\, \mathrm{CNG}\;.\\
\end{cases}
\ee
where $\mathrm{K}\equiv k/(aH)$ and $k$ is the wavenumber of the perturbation, $h$ and $\eta$ are the scalar metric 
perturbations in the SG, and where a prime denotes a derivative with respect to $\ln{a}$. Since the SG is defined as 
the rest frame of the CDM fluid, we see that $T$ is nothing else than the velocity perturbation of the CDM fluid 
evaluated in the CNG.

To work in a gauge invariant way, with respect to the synchronous and conformal Newtonian gauges, we can define a gauge 
invariant velocity perturbation as
\be\label{eq:T}
 \hat{\Theta} \equiv \Theta + 3(1+w)T\;,
\ee
In the same line of thought, using the variable $T$, the evolution equations for the gauge invariant density 
perturbation and rescaled velocity perturbation can be written in a way that is valid for both gauges 
\citep{Battye2016a}. These are the so-called \textit{perturbed fluid equations} and are given by

\begin{eqnarray}\label{eq:pfe}
 \Delta^{\prime} - 3w\Delta - 2\Pi + \qsubrm{g}{K}\qsubrm{\epsilon}{H}\hat{\Theta} & = & 3(1+w)X\;, 
 \label{eq:deltadot}\nonumber \\
 \hat{\Theta}^\prime + 3\left(\qsubrm{c}{a}^2-w+\tfrac{1}{3}\qsubrm{\epsilon}{H}\right)\hat{\Theta} - 
 3\qsubrm{c}{a}^2\Delta - 2\Pi - 3\Gamma & = & 3(1+w)Y\;, \nonumber\label{eq:Thetadot}\\
\end{eqnarray}
where $\qsubrm{c}{a}^2\equiv dP/d\rho$ is the adiabatic sound speed and 
$\qsubrm{g}{K}\equiv 1+\rm{K}^2/(3\qsubrm{\epsilon}{H})$, with $\qsubrm{\epsilon}{H}\equiv -H^\prime/H$ and where
\bse
 \bea
  X & \equiv & \begin{cases}
   \eta^\prime + \qsubrm{\epsilon}{H}T & \mathrm{in}\,\,\mathrm{the}\,\,\mathrm{SG}\;, \\
   \phi^\prime + \psi & \mathrm{in}\,\,\mathrm{the}\,\, \mathrm{CNG}\;,\\
  \end{cases}\\
  Y & \equiv & \begin{cases}
   T^\prime+\qsubrm{\epsilon}{H}T & \mathrm{in}\,\,\mathrm{the}\,\,\mathrm{SG}\;,\\
   \psi & \mathrm{in}\,\,\mathrm{the}\,\,\mathrm{CNG}\;.\\
  \end{cases}
 \eea
\ese
Finally, $\Pi$ is the \textit{perturbed scalar anisotropic stress}
\footnote{\label{fn:MB}Note that our $\theta$ and $\Pi$ differ from $\qsubrm{\theta}{MB}$ and $\qsubrm{\sigma}{MB}$ 
(anisotropic stress) as defined in \cite{Ma1995}, by $\qsubrm{\theta}{MB}=\frac{k^2}{a}\theta$ and 
$(\rho+P)\qsubrm{\sigma}{MB}=-\tfrac{2}{3}\rho\Pi$.} 
and $\Gamma$ is the gauge invariant \textit{entropy perturbation}. The gauge invariant entropy perturbation can be 
expressed in terms of the perturbed pressure, density and rescaled velocity as
\be
 \Gamma = \frac{\delta P}{\rho} - \qsubrm{c}{a}^2(\Delta - \Theta)\;.
\ee
The perturbed fluid equations \eqref{eq:pfe} are valid for both matter (that we shall denote with a subscript `m') and 
dark energy (that we shall denote with subscript `de') fluid variables.

The Einstein-Boltzmann code \verb|CLASS| \citep{Lesgourgues2011a,Blas2011} written in C provides the infrastructure 
required to solve the dynamics of matter perturbations. We have incorporated the EoS approach for dark energy 
perturbations into \verb|CLASS| and dubbed the modified code \verb|CLASS_EOS_FR|. The code is publicly available on the 
internet \footnote{website:\href{https://github.com/borisbolliet/class_eos_fr_public}
{https://github.com/borisbolliet/class\_eos\_fr\_public}}. We have implemented the perturbed fluid equations 
\eqref{eq:pfe} for dark energy perturbations in this exact same form. We now describe the remaining technical steps 
necessary to close the system of equation \eqref{eq:pfe} and integrate it in the code.

As prescribed by the EoS approach, we expand the perturbed dark energy anisotropic stress and gauge invariant entropy 
perturbation in terms of the perturbed fluid variables. These are the so-called \textit{equations of state} for dark 
energy perturbations and are written as
\begin{widetext}
 \begin{align}
  \begin{array}{ccccccccccc}
   \Pi_{{\scriptscriptstyle\mathrm{de}}} & = & 
     c_{{\scriptscriptstyle\mathrm{\Pi\Delta_{{\scriptscriptstyle\mathrm{de}}}}}}
     \Delta_{{\scriptscriptstyle \mathrm{de}}} &+& 
     c_{{\scriptscriptstyle \mathrm{\Pi\Theta_{{\scriptscriptstyle \mathrm{de}}}}}}
     {\hat{\Theta}}_{{\scriptscriptstyle \mathrm{de}}} &+& 
     c_{{\scriptscriptstyle \mathrm{\Pi\Delta_{{\scriptscriptstyle \mathrm{m}}}}}}
     \Delta_{{\scriptscriptstyle \mathrm{m}}} &+& 
     c_{{\scriptscriptstyle \mathrm{\Pi\Theta_{{\scriptscriptstyle \mathrm{m}}}}}}
     {\hat{\Theta}}_{{\scriptscriptstyle \mathrm{m}}} &+& 
     c_{{\scriptscriptstyle \mathrm{\Pi\Pi_{{\scriptscriptstyle \mathrm{m}}}}}}
     {\Pi}_{{\scriptscriptstyle \mathrm{m}}}\;,\\
   \Gamma_{{\scriptscriptstyle \mathrm{de}}} & = & 
     c_{{\scriptscriptstyle \mathrm{\Gamma\Delta_{{\scriptscriptstyle \mathrm{de}}}}}}
     \Delta_{{\scriptscriptstyle \mathrm{de}}} &+& 
     c_{{\scriptscriptstyle \mathrm{\Gamma\Theta_{{\scriptscriptstyle \mathrm{de}}}}}}
     {\hat{\Theta}}_{{\scriptscriptstyle \mathrm{de}}} &+& 
     c_{{\scriptscriptstyle \mathrm{\Gamma\Delta_{{\scriptscriptstyle \mathrm{m}}}}}}
     \Delta_{{\scriptscriptstyle \mathrm{m}}} &+& 
     c_{{\scriptscriptstyle \mathrm{\Gamma\Theta_{{\scriptscriptstyle \mathrm{m}}}}}}
     {\hat{\Theta}}_{{\scriptscriptstyle \mathrm{m}}}&+& 
     c_{{\scriptscriptstyle \mathrm{\Gamma\Gamma_{{\scriptscriptstyle \mathrm{m}}}}}}
     {\Gamma}_{{\scriptscriptstyle \mathrm{m}}}\;,
  \end{array}\label{eq:eos}
 \end{align}
\end{widetext}
where the coefficients $c_{\alpha\beta}$ are \textit{a priori} scale and time dependent functions, but shall only 
depend on the homogeneous background quantities, such as the Hubble parameter, the background equation of state of dark 
energy, or the adiabatic sound speeds. These functions are specified for each dark energy and modified gravity model, 
e.g., see \citep{Battye2016a} for $f(\mathcal{R})$ gravity and \citep{Battye2017} for Generalised Einstein-Aether. 
Note that the equations of state for perturbations for generic $f(\mathcal{R})$ models can also be obtained starting 
from a general Horndeski model and specifying the appropriate free functions to match with $f(\mathcal{R})$ theories. 
In this case, the expressions for the coefficients of $c_{\alpha\beta}$ are as reported in appendix~\ref{sect:EoS_EFT}.

Initial conditions for dark sector perturbations are set at an early time, $\qsubrm{a}{ini}$, when dark energy is 
subdominant, i.e., $\qsubrm{\Omega}{de}(\qsubrm{a}{ini})\ll 1$ where $\qsubrm{\Omega}{de}$ is the dark energy density 
parameter. If not specified from the specific dark energy model, appropriate initial conditions for the dark energy 
perturbations are generally: $\qsubrm{\Delta}{de}(\qsubrm{a}{ini}) = \qsubrm{\Theta}{de}(\qsubrm{a}{ini}) = 0$. Note 
that when there exists an attractor for the dark energy perturbations during matter domination, it is numerically more 
efficient to set initial conditions that match the attractor (see Sec.~\ref{sect:results}).

In order to evaluate the equation of state \eqref{eq:eos} and integrate equations \eqref{eq:pfe}, we collect the 
perturbed matter fluid variables at every time step. In our code, we do this in the following way. First, we obtain the 
total matter gauge invariant density perturbation via
\bea
 \qsubrm{\Omega}{m}{\Delta}_{\scriptscriptstyle\mathrm{m}} & = -\tfrac{2}{3}\mathrm{K}^2 Z - 
 \qsubrm{\Omega}{de}{\Delta}_{\scriptscriptstyle\mathrm{de}} \label{eq:dm}
 \,\,\mathrm{with} \nonumber\\
 Z & \equiv \begin{cases}
  \eta - T & \mathrm{in}\,\,\mathrm{the}\,\,\mathrm{SG}\\ 
  \phi & \mathrm{in}\,\,\mathrm{the}\,\, \mathrm{CNG} \\
 \end{cases}
\eea
and the gauge invariant matter velocity perturbation via 
$\qsubrm{\Omega}{m}{\hat{\Theta}}_{\scriptscriptstyle\mathrm{m}} = 
2X-\qsubrm{\Omega}{de}{\hat{\Theta}}_{\scriptscriptstyle\mathrm{de}}$, see \citep{Battye2016a} where these equations 
are derived. Next, the matter pressure perturbation $\qsubrm{\delta P}{m}$ and the matter anisotropic stress 
$\sigma_{{\scriptscriptstyle{\textrm{m}}}}^{{\scriptscriptstyle{\textrm{class}}}}$ are available in \verb|CLASS|. 
We use them to compute the matter anisotropic stress perturbation (in our convention) $\qsubrm{\Pi}{m}$ and the matter 
gauge invariant entropy perturbation as
\bea
 \qsubrm{\rho}{m}\qsubrm{\Pi}{m}  &=&  
 -\tfrac{3}{2}\left\langle(\qsubrm{\rho}{m}+\qsubrm{P}{m})
 {\sigma_{{\scriptscriptstyle{\textrm{m}}}}^{{\scriptscriptstyle{\textrm{class}}}}}\right\rangle,\\
 \qsubrm{\rho}{m} \qsubrm{\Gamma}{m}  &=&  \left\langle\qsubrm{\delta P}{m}\right\rangle - 
 c_{a,\scriptscriptstyle{\mathrm{m}}}^2(\qsubrm{\Delta}{m} - \qsubrm{\Theta}{m})\;,
\eea
where the brackets mean a sum over \textit{all} the matter fluid components, i.e., baryons, CDM, photons and neutrinos, 
and
\be
 c_{a,\scriptscriptstyle{\mathrm{m}}}^2 = 
 \frac{w_{\scriptscriptstyle{\mathrm{m}}}\Omega_{\scriptscriptstyle{\mathrm{m}}} + 
 \left\langle w_{\scriptscriptstyle{\mathrm{m}}}^{2}\Omega_{\scriptscriptstyle{\mathrm{m}}}\right\rangle}
 {\left(1+w_{\scriptscriptstyle{\mathrm{m}}}\right)\Omega_{\scriptscriptstyle{\mathrm{m}}}}\;,
\ee
is the matter adiabatic sound speed, where $\qsubrm{\Omega}{m} \equiv 1-\qsubrm{\Omega}{de}$ and 
$\qsubrm{w}{m} \equiv \left\langle \qsubrm{w}{m} \qsubrm{\Omega}{m} \right\rangle/\qsubrm{\Omega}{m}$ are the matter 
density parameter and background equation of state respectively. Last, we update the total stress energy tensor 
accordingly as
\bea
 \qsubrm{\delta\rho}{tot} & = & \left\langle \qsubrm{\delta\rho}{m}\right\rangle + 
 \qsubrm{\rho}{de}\qsubrm{\Delta}{de}- 
 \qsubrm{\rho}{de}\qsubrm{\Theta}{de}\nonumber\\
 (\qsubrm{\rho}{tot} + \qsubrm{P}{tot})
 {\theta_{{\scriptscriptstyle{\textrm{tot}}}}^{{\scriptscriptstyle{\textrm{class}}}}} & = & 
 \left\langle(\qsubrm{\rho}{m}+\qsubrm{P}{m})
 {\theta_{{\scriptscriptstyle{\textrm{m}}}}^{{\scriptscriptstyle{\textrm{class}}}}}\right\rangle + 
 \tfrac{1}{3}\mathrm{K}^2 aH\qsubrm{\rho}{de}\qsubrm{\Theta}{de}\nonumber\\
 (\qsubrm{\rho}{tot}+\qsubrm{P}{tot})
 {\sigma_{{\scriptscriptstyle{\textrm{tot}}}}^{{\scriptscriptstyle{\textrm{class}}}}} & = & 
 \left\langle(\qsubrm{\rho}{m}+\qsubrm{P}{m})
 {\sigma_{{\scriptscriptstyle{\textrm{m}}}}^{{\scriptscriptstyle{\textrm{class}}}}}\right\rangle - 
 \tfrac{2}{3}\qsubrm{\rho}{de}\qsubrm{\Pi}{de}\nonumber\\
 \qsubrm{\delta P}{tot} & = & \left\langle \qsubrm{\delta P}{m}\right\rangle +\qsubrm{\rho}{de} \qsubrm{\Gamma}{de} + 
 c_{a,\scriptscriptstyle{\mathrm{de}}}^2 \qsubrm{\rho}{de}(\qsubrm{\Delta}{de} - \qsubrm{\Theta}{de})\;.\nonumber
\eea
See footnote \ref{fn:MB} for the \verb|CLASS| perturbed velocity, which follows the conventions of \cite{Ma1995}.

Although the numerical integration can be carried out either in the conformal Newtonian gauge or in the synchronous 
gauge in \verb|CLASS_EOS_FR|, we find that, in the super-Hubble regime, i.e., $\mathrm{K}^2\ll1$, the synchronous gauge 
performs better than the conformal Newtonian gauge.

\section{A brief reminder on the designer \texorpdfstring{$f(\mathcal{R})$}{f(R)} gravity models}\label{sect:fR}
In $f(\mathcal R)$ gravity, the $f(\mathcal R)$ functions are solutions to a second order differential equation given 
by the projection of the stress-energy tensor of $f(\mathcal{R})$ on the time direction, which can be written as 
\citep{Song2007,Pogosian2008,Nojiri2009,Dunsby2010,Lombriser2012}
\be\label{eqn:fR}
 f^{\prime\prime} + 
 \left(3\epsilon_{{\scriptscriptstyle \mathrm{H}}}-1-
 \frac{\bar{\epsilon}_{{\scriptscriptstyle\mathrm{H}}}^{\prime}}
 {\bar{\epsilon}_{{\scriptscriptstyle\mathrm{H}}}}\right) f^{\prime} - 
 \bar{\epsilon}_{{\scriptscriptstyle\mathrm{H}}}f = 
 6H^{2}\bar{\epsilon}_{{\scriptscriptstyle\mathrm{H}}}\Omega_{{\scriptscriptstyle \mathrm{de}}}\;,
\ee
where the prime still denotes a derivative with respect to $\ln{a}$ and 
$\bar{\epsilon}_{{\scriptscriptstyle \mathrm{H}}} = 
{\epsilon}_{{\scriptscriptstyle\mathrm{H}}}^\prime + 4\epsilon_{{\scriptscriptstyle \mathrm{H}}} - 
2\epsilon_{{\scriptscriptstyle\mathrm{H}}}^{2}$ (see Eq. (2.6a) of \citep{Battye2016a} for the derivation in our 
conventions). This equation holds for any $f(\mathcal{R})$ gravity model and at any time during the expansion history.
\begin{figure}
 \begin{center}
  \includegraphics[width=7.5cm,angle=0]{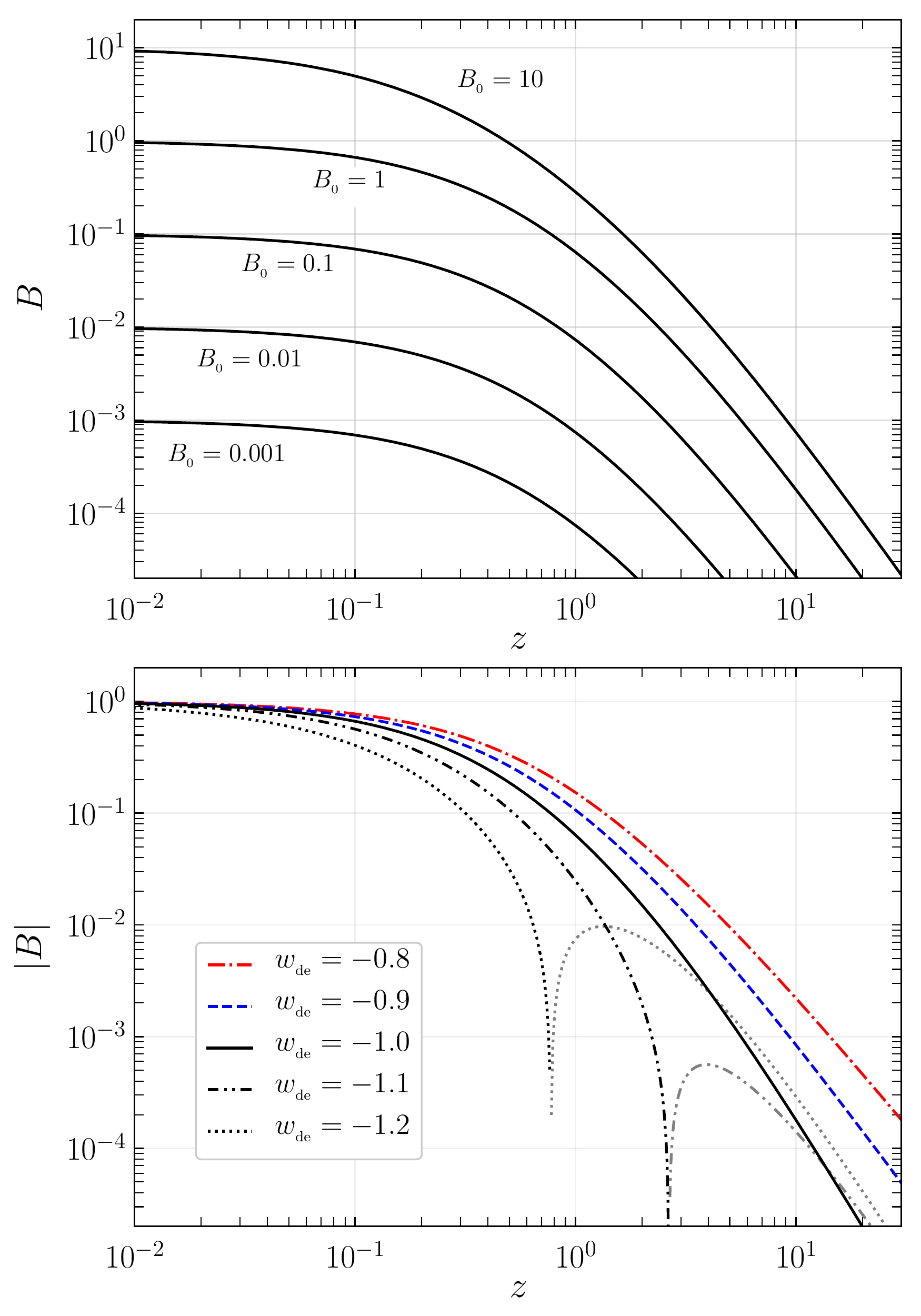}
  \caption[justified]{The redshift evolution of 
  $B=-({f_{\mathcal{R}}^{\prime}}/[{\qsubrm{\epsilon}{H}(1+f_{\mathcal{R}})}])$ for different designer $f(\mathcal{R})$ 
  models. Unless otherwise written, we chose $\qsubrm{w}{de}=-1$ and $\qsubrm{B}{0}=1$. A grey line indicates negative 
  values. The background cosmology was set to $h=0.7$, $\qsubrm{\Omega}{de}=0.7$ and $\qsubrm{\Omega}{b}h^2=0.022$, 
  where $h=H_0/100$ is the reduced Hubble parameter.}
  \label{fig:bgfr}
 \end{center}
\end{figure}
During the non-relativistic matter era, i.e., $\qsubrm{w}{m}=0$, this equation simplifies because 
$\qsubrm{\epsilon}{H} = 3/2$, $\qsubrm{\bar{\epsilon}}{H}^\prime = 0$ and 
$\qsubrm{\bar{\epsilon}}{H} = \qsubrm{\epsilon}{H} = 3/2$ (see Eq. (2.5) of \citep{Battye2016a}). In this regime, the 
solutions to \eqref{eqn:fR} are
\be\label{eq:fsol}
 f(a) = C\left\{{b}_{+}a^{n_{+}}+{b}_{-}a^{n_{-}}+e^{-\int 3(1+\qsubrm{w}{de})\mathrm{d}\ln a}\right\}\;,
\ee
with $n_{\pm} = \tfrac{7}{4}(-1\pm\sqrt{73/49})$ and 
$C = \frac{6\Omega_{{\scriptscriptstyle\mathrm{de}}}^{{\scriptscriptstyle 0}}H_{0}^{2}}
{6\qsubrm{w}{de}^{2}+5\qsubrm{w}{de}-2}$. 
Solutions with ${b}_{-}\neq 0$ are not admissible because they break the condition 
$\lim_{a\rightarrow 0}f_{\mathcal{R}}=0$ \citep{Khoury2004a,Khoury2004b,Hu2007}, where a subscript `$\mathcal{R}$' 
means a derivative with respect to the Ricci scalar. We conclude that any viable $f(\mathcal{R})$ gravity model can be 
parameterized, in the non-relativistic matter era, by the \textit{a priori} time dependent equation of state 
$\qsubrm{w}{de}(a)$ and a constant number $b_+$. We then trade $b_+$ for the more commonly used parameter
\be\label{eq:bo}
 B \equiv -\frac{{f_{\mathcal{R}}^{\prime}}}{\qsubrm{\epsilon}{H}(1+f_{\mathcal{R}})}\;,
\ee
evaluated today and dubbed $\qsubrm{B}{0}$, since there is a one-to-one correspondence between $b_{+}$ and 
$\qsubrm{B}{0}$. 
From here, there are two ways to proceed. The first possibility is to specify explicitly a $f(\mathcal{R})$ function at 
all time, and then extract the time evolution of $\qsubrm{\Omega}{de}$ and $\qsubrm{w}{de}$ from the time derivatives 
of $f$. 
The second possibility is to specify a time evolution for $\qsubrm{\Omega}{de}$ and $\qsubrm{w}{de}$ and then integrate 
Eq.~(\ref{eqn:fR}) to get $f(\mathcal{R})$ at all time. This latter approach is the so-called designer, or mimetic, 
$f(\mathcal{R})$ approach and leads to the $f(\mathcal{R})$ gravity models that we are interested in. Designer models 
are particularly interesting because their functional form is dictated by the chosen background evolution of the 
dark fluid and therefore there is no arbitrariness in how the $f(\mathcal{R})$ Lagrangian looks like. In this way the 
wanted background evolution is achieved exactly and the model has less degrees of freedom: the only value to be 
determined is $\qsubrm{B}{0}$, which ultimately will dictate the strength of the perturbations.

In \verb|CLASS_EOS_FR|, we have implemented the designer models with constant equation of state $\qsubrm{w}{de}$. The 
user specifies a value for $\qsubrm{w}{de}$ and $\qsubrm{B}{0}$, then the code explores a range of $b_{+}$ solving 
\eqref{eqn:fR}, between $\qsubrm{a}{ini}$ and today, until it finds the value of $b_{+}$ that leads to the desired 
value of $\qsubrm{B}{0}$. Note that the solution in \eqref{eq:fsol} is singular for $\qsubrm{w}{de}\simeq0.30$ and 
$\qsubrm{w}{de}\simeq-1.13$, however as long as one avoids the two poles, the numerical integration is efficient.

In \citep{Song2007,Battye2016a}, the designer models with $\qsubrm{w}{de}=-1$ were studied at both the 
background and perturbation levels. Here, we consider as well the designer models with $\qsubrm{w}{de}\neq-1$ (and 
$\qsubrm{w}{de}^\prime=0$), i.e., the ones that mimic a $w$CDM expansion history.

In Fig.~\ref{fig:bgfr} we show the redshift evolution of a set of solutions to \eqref{eqn:fR} for different values 
of $\qsubrm{B}{0}$ and $\qsubrm{w}{de}$. We present $B$, rather than $f(\mathcal{R})$ itself, because this is the main 
quantity entering the equations of state for perturbation $\qsubrm{\Pi}{de}$ and $\qsubrm{\Gamma}{de}$ 
\cite{Battye2016a}. On the bottom panel we fix $\qsubrm{B}{0}=1$ and vary $\qsubrm{w}{de}$. For models with 
$\qsubrm{w}{de}<-1$, $B$ starts being negative and eventually becomes positive at late time. This can be described 
analytically with Eq. \eqref{eq:fsol}, see, e.g., \citep{Song2007}. On the top panel we fix $\qsubrm{w}{de}=-1$ and 
vary $\qsubrm{B}{0}$. As can be seen, as soon as dark energy dominates, i.e., $z\lesssim 0.3$, $B$ settles to its final 
value $\qsubrm{B}{0}$. Changing the value of $\qsubrm{w}{de}$ essentially amounts to a shift of the curves on this plot 
because for a less negative $\qsubrm{w}{de}$ dark energy dominates earlier. The bottom panel shows that when we keep 
$\qsubrm{B}{0}$ fixed, $B$ grows more slowly for less negative $\qsubrm{w}{de}$. More precisely, with Eq. 
\eqref{eq:fsol} in the matter era, one finds $B\sim z^{3\qsubrm{w}{de}}$.

\section{Evolution of perturbations in the dark energy fluid of \texorpdfstring{$f(\mathcal{R})$}{f(R)} gravity}
\label{sect:results}
In this section we investigate numerically and analytically the evolution of cosmological perturbations for the 
designer $f(\mathcal{R})$ gravity models described in Sec.~\ref{sect:fR}. To this aim, we use the formalism of the 
EoS approach described in Sec.~\ref{sect:EoS}.

\begin{figure}
 \begin{center}
  \includegraphics[width=7cm,angle=0]{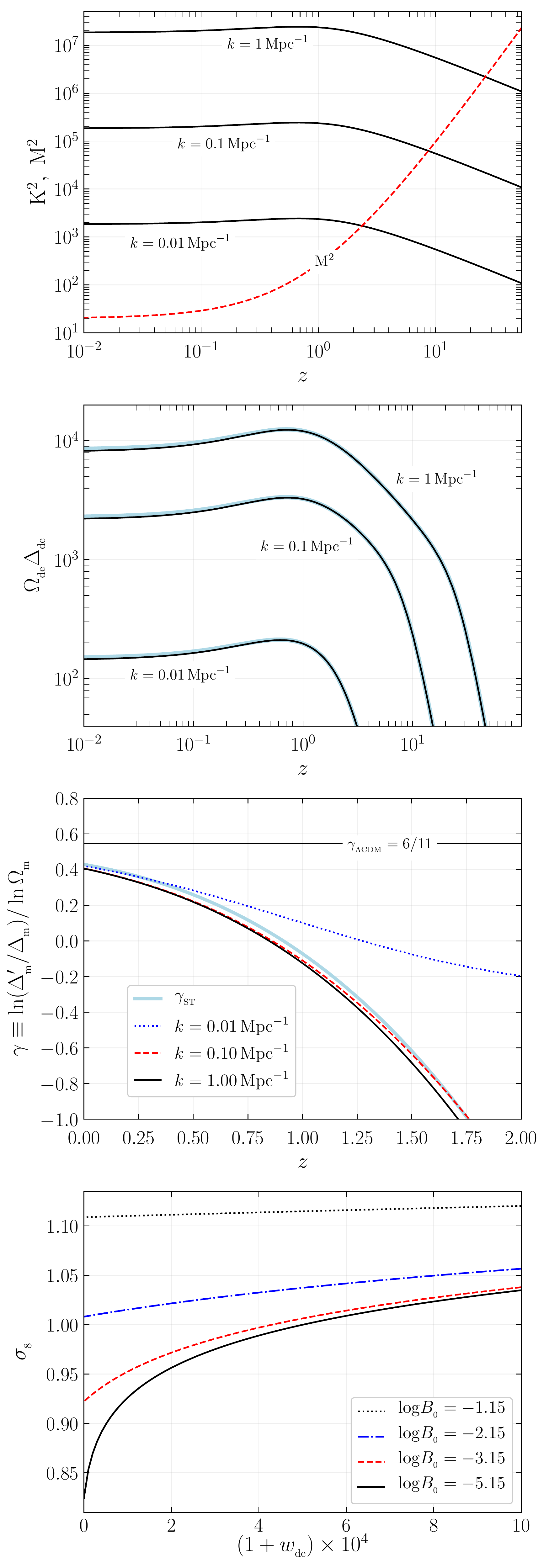}
  \caption[justified]{The redshift evolution of $\mathrm{K}^2$ for three wavenumbers and $\mathrm{M}^2$ (dashed line in 
  the top panel), $\qsubrm{\Omega}{de}\qsubrm{\Delta}{de}$ and $\gamma$ (middle panels) and $\qsubrm{\sigma}{8}$ as a 
  function of $\qsubrm{w}{de}$ (bottom panel) for different designer $f(\mathcal{R})$ models. The attractor solution 
  \eqref{eq:dde} and the growth index $\qsubrm{\gamma}{ST}$ \eqref{eqn:gammaST} are the thick grey lines. Unless 
  otherwise written, we chose $\qsubrm{w}{de}=-1$ and $\qsubrm{B}{0}=0.1$ as well as the same cosmology as in 
  Fig.~\ref{fig:bgfr} with $A_s=2.2\times 10^{-9}$ and $n_s=0.96$.}
  \label{fig:deltade}
 \end{center}
\end{figure}

To gain some understanding about the behaviour of the cosmological perturbations, we consider the expressions of the 
equations of state for perturbations for a $f(\mathcal{R})$ fluid with constant equation of state parameter, i.e., 
$c_{a,\scriptscriptstyle{\mathrm{de}}}^2 = \qsubrm{w}{de}$, and when the matter sector is dominated by non-relativistic 
species, i.e., $\qsubrm{w}{m} = \qsubrm{\Pi}{m} = \qsubrm{\Gamma}{m} =0$, as is the case after radiation domination. 
Furthermore, we focus on modes that enter the Hubble horizon before dark energy dominates so that we have 
$\mathrm{K}^2\gg1$ at all time. This assumption holds for wavenumbers in the observational range of interest to us 
(see top panel of Fig.~\ref{fig:deltade}). Finally we assume $B\ll1$, which is true at all times if 
$\qsubrm{B}{0}\ll1$ and is equivalent to $\mathrm{M^2}\gg1$, with 
$\mathrm{M}^2\equiv 2\qsubrm{\bar{\epsilon}}{H}/(\qsubrm{\epsilon}{H}B)$.
In this regime, the equations of state for dark energy perturbations simplify to
\bse
 \bea
  \qsubrm{\Pi}{de} & = & \qsubrm{\Delta}{de}\;,\label{eqn:Pi_de}\\
  \qsubrm{\Gamma}{de} & = & \left\{\tfrac{1}{3}-\qsubrm{w}{de}+\tfrac{\mathrm{M}^2}{\mathrm{K}^2}\right\}
  \qsubrm{\Delta}{de} + \tfrac{1}{3}\tfrac{\qsubrm{\Omega}{m}}{\qsubrm{\Omega}{de}}\qsubrm{\Delta}{m}\,.
 \eea
\ese

Using the field equation (3.11a) and (3.11b) in \citep{Battye2016a}, the perturbed fluid equations \eqref{eq:pfe} can 
be rewritten as a system of two coupled second order differential equations for the gauge invariant density 
perturbations,
\bse\label{eqn:Deltapp}
 \bea
  \qsubrm{\Delta}{m}^{\prime\prime} + (2-\qsubrm{\epsilon}{H})\qsubrm{\Delta}{m}^\prime - 
  \tfrac{3}{2}\qsubrm{\Omega}{m}\qsubrm{\Delta}{m} & = & -\tfrac{3}{2}\qsubrm{\Omega}{de}\qsubrm{\Delta}{de}\;,
  \label{eq:dmde}\\
  \qsubrm{\Delta}{de}^{\prime\prime} + (2-\qsubrm{\epsilon}{H})\qsubrm{\Delta}{de}^\prime + 
  ({\mathrm{K}^2}+{\mathrm{M}^2})\qsubrm{\Delta}{de} & = & -\tfrac{1}{3}\tfrac{\qsubrm{\Omega}{m}}{\qsubrm{\Omega}{de}}
  \mathrm{K}^2\qsubrm{\Delta}{m}\,.\,\,\,\,\,\,\quad
  \label{eq:dede}
 \eea
\ese
For the modes of interest, this set of equations provides a faithful description of the dynamics of cosmological 
perturbations as long as $B\ll 1$. Again, this is always the case before dark energy dominates (irrespective of 
$\qsubrm{B}{0}$). In addition if $\qsubrm{B}{0}\ll 1$, then these equations are also valid during dark energy 
domination, because $B$ is always smaller than $\qsubrm{B}{0}$ (see Fig.~\ref{fig:bgfr}). Let us assume 
$\qsubrm{B}{0}=\mathcal{O}(1)$, or equivalently $\mathrm{M}^2 \gg 1$, from now on. As we shall see in 
Sec.~\ref{sect:obs}, this is a reasonable assumption given current observational constraints.

The differential equation \eqref{eq:dede} for the gauge invariant energy density perturbation is similar to an harmonic 
oscillator with a time dependent frequency $\omega^2=\mathrm{K}^2+\mathrm{M}^2\gg 1$. Since the oscillatory time scale 
is much smaller than the damping time scale. i.e., the expansion rate, the homogeneous solution to \eqref{eq:dede} 
becomes rapidly subdominant compared to the particular solution. This confirms that the specific values for the initial 
dark energy perturbations are not important. More precisely, the dark energy density perturbation relates to the matter 
density perturbation via
\be\label{eq:dde}
 \qsubrm{\Omega}{de}\qsubrm{\Delta}{de} = 
 -\frac{1}{3}\frac{\mathrm{K}^2}{\mathrm{K}^2+\mathrm{M}^2}\qsubrm{\Omega}{m}\qsubrm{\Delta}{m}\;.
\ee
We refer to \citep{Tsujikawa2009} for the same result formulated in a different language. In our code, we set the 
initial conditions for $\qsubrm{\Delta}{de}$ and $\qsubrm{\Theta}{de}$ according to \eqref{eq:dde} at a time such that 
$\mathrm{K}^2/[3(\mathrm{K}^2+\mathrm{M}^2)]=
|\qsubrm{\Omega}{de}\qsubrm{\Delta}{de}/\qsubrm{\Omega}{m}\qsubrm{\Delta}{m}|=0.01$. 
Note that given this criterion, the initial starting time for dark energy perturbation depends on the wavenumber.

We deduce from \eqref{eq:dde} the two regimes for the behaviour of sub-horizon modes: (i) the general relativistic (GR) 
regime when $\mathrm{K}^2\ll\mathrm{M}^2$, i.e., at early time, and (ii) the scalar-tensor (ST) regime when 
$\mathrm{K}^2\gg\mathrm{M}^2$, i.e., at late time. This implies $\qsubrm{\Omega}{de}\qsubrm{\Delta}{de} = 
-\frac{\mathrm{K}^2}{\mathrm{M}^2}\qsubrm{\Omega}{m}\qsubrm{\Delta}{m}$ in the GR regime, and 
$\qsubrm{\Omega}{de}\qsubrm{\Delta}{de} = -\frac{1}{3}\qsubrm{\Omega}{m}\qsubrm{\Delta}{m}$ in the ST regime. 
Moreover, in both regimes, the differential equation for the matter perturbation \eqref{eq:dmde} becomes
\be\label{eqn:gfe}
 \qsubrm{\Delta}{m}^{\prime\prime} + (2-\qsubrm{\epsilon}{H})\qsubrm{\Delta}{m}^\prime - 
 \tfrac{3}{2}\varepsilon\qsubrm{\Omega}{m}\qsubrm{\Delta}{m} = 0\;,
\ee
where $\varepsilon\equiv (4\mathrm{K}^2+3\mathrm{M}^2)/(3\mathrm{K}^2+3\mathrm{M}^2)$ can be interpreted as a 
\textit{modification} to the gravitational constant \citep{Starobinsky2007}. One has $\varepsilon=4/3$ in the ST regime 
and $\varepsilon=1$ in the GR regime. Since one has $\mathrm{K}^2\sim z^{-1}$ and 
$\mathrm{M}^2\sim z^{-3\qsubrm{w}{de}}$ during the matter era, the ST regime starts earlier for less negative 
$\qsubrm{w}{de}$.

Eq. \eqref{eq:dde} and \eqref{eqn:gfe} enable a clear discussion of the dynamics of cosmological perturbations in 
$f(\mathcal{R})$ gravity. Before doing so, we go one step further and obtain the growth index 
$\gamma\equiv\ln\mathrm{f}/\ln\qsubrm{\Omega}{m}$ of the matter perturbation \citep{Peebles1980}, where 
$\mathrm{f}\equiv\qsubrm{\Delta}{m}^\prime/\qsubrm{\Delta}{m}$ is the growth rate.

Taking the time derivative of the growth rate and using \eqref{eqn:gfe} we find
\be\label{eqn:ge}
 \gamma^{\prime}+ \frac{3\qsubrm{w}{de}\qsubrm{\Omega}{de}}{\ln{\qsubrm{\Omega}{m}}}\gamma+
 \frac{\qsubrm{\Omega}{m}^\gamma}{\ln{\qsubrm{\Omega}{m}}}-
 \frac{3\qsubrm{\Omega}{m}^{1-\gamma}}{2\ln{\qsubrm{\Omega}{m}}}\varepsilon=
 \frac{3\qsubrm{w}{de}\qsubrm{\Omega}{de}-1}{2\ln{\qsubrm{\Omega}{m}}}\;,
\ee
for the growth index. To linearise this equation, we use the approximations 
$\ln \qsubrm{\Omega}{m}\approx-\qsubrm{\Omega}{de}$ and $\qsubrm{\Omega}{m}^\gamma\approx1-\gamma\qsubrm{\Omega}{de}$ 
which are valid when $\qsubrm{\Omega}{de}=\mathcal{O}(1)$. We get
\be\label{eq:gp}
 \gamma^{\prime} + 
 \left(1-3\qsubrm{w}{de}+\tfrac{3}{2}\varepsilon\right)\gamma = 
 \tfrac{3}{2}\left(\tfrac{1-\varepsilon}{\qsubrm{\Omega}{de}}+\varepsilon-\qsubrm{w}{de}\right)\;.
\ee
This can be solved analytically for a constant $\varepsilon$. We find 
\be\label{eqn:gamma}
 \gamma = \frac{3(1-\varepsilon)}{2+3\varepsilon}\frac{\qsubrm{\Omega}{m,0}}{\qsubrm{\Omega}{de,0}}
          (1+z)^{-3\qsubrm{w}{de}}+\frac{3(\varepsilon-\qsubrm{w}{de})}{2+3\varepsilon-6\qsubrm{w}{de}}\;.
\ee
In the GR regime the first term on the right hand side vanishes, and the second term gives a constant 
$\gamma_{_{w\mathrm{CDM}}}=3(1-\qsubrm{w}{de})/(5-6\qsubrm{w}{de})$, i.e., the $w$CDM growth index. If in addition 
$\qsubrm{w}{de}=-1$, the the growth index is $\gamma_{_{\Lambda\mathrm{CDM}}}=6/11\approx0.545$, i.e., the well-known 
$\Lambda$CDM result. In the ST regime, the growth index is not constant any more due to the first term on the right 
hand side. We find
\be\label{eqn:gammaST}
 \gamma_{_\mathrm{ST}} = \frac{1}{2} + \frac{1}{6(1-\qsubrm{w}{de})} 
                         -\frac{\qsubrm{\Omega}{m,0}}{6\qsubrm{\Omega}{de,0}}(1+z)^{-3\qsubrm{w}{de}}\;.
\ee
Since the first term on the right hand side of \eqref{eqn:gamma} is always negative, we have 
$\gamma_{_\mathrm{ST}}<\gamma_{_{w\mathrm{CDM}}}$ as well as $\gamma_{_\mathrm{ST}}<\gamma_{_{\Lambda\mathrm{CDM}}}$.

We now summarise the important consequences for the dynamics of perturbations in $f(\mathcal{R})$ gravity that are 
deduced from the above considerations.

\begin{enumerate}
\item For $B<0$ (or $\mathrm{M}^2<0$), the homogeneous solution to \eqref{eq:dede} is unstable. Therefore, the gauge 
invariant density perturbation for both matter and dark energy grows exponentially with time. This is not compatible 
with the dynamics of matter perturbations in the matter dominated era, and consequently $f(\mathcal{R})$ models with 
$\qsubrm{B}{0}<0$ or $\qsubrm{w}{de}<-1$ are not viable, see Fig.~\ref{fig:bgfr}.
\item \label{point2} The gauge invariant density perturbation in the dark energy component relates to that of the 
matter component in a simple way given in \eqref{eq:dde}. In the GR regime, the dark energy perturbation is negligible 
compared to the matter perturbation, while in the ST regime both are of the same magnitude, see Fig.~\ref{fig:deltade}.
\item \label{point3} In $w$CDM, for less negative $\qsubrm{w}{de}$ structures are less gravitationally bounded compared 
to $\Lambda$CDM because dark energy starts dominating earlier. 
Hence there is an \textit{anti-correlation} between $\qsubrm{w}{de}$ and the amplitude of clustering, i.e. 
$\qsubrm{\sigma}{8}$, in $w$CDM models (see, e.g. Fig.~16 of \citep{Komatsu2009}). In $f(\mathcal{R})$ gravity matter 
perturbations grow at a faster rate than in $w$CDM and $\Lambda$CDM because 
$\gamma_{_\mathrm{ST}}<\gamma_{_{w\mathrm{CDM}}}$ [see Eq. \eqref{eqn:gammaST}]. This, combined with the fact that the 
ST regime starts earlier for less negative $\qsubrm{w}{de}$, implies a \textit{correlation} between $\qsubrm{w}{de}$ 
and $\qsubrm{\sigma}{8}$ (see bottom panel of Fig.~\ref{fig:deltade}), and can be used to discriminate between 
$f(\mathcal{R})$ gravity and $w$CDM models of dark energy [see also the next section for a comparison between 
$w$CDM and $f(\mathcal{R})$ models].
\end{enumerate}
In the next section we compute relevant observables that we use to set observational constraints on the designer 
$f(\mathcal{R})$ gravity models.

\section{Impact of \texorpdfstring{$f(\mathcal{R})$}{f(R)} gravity on observables and constraints}\label{sect:obs}

\begin{figure*}
 \begin{center}
  \includegraphics[height=8.5cm,angle=0]{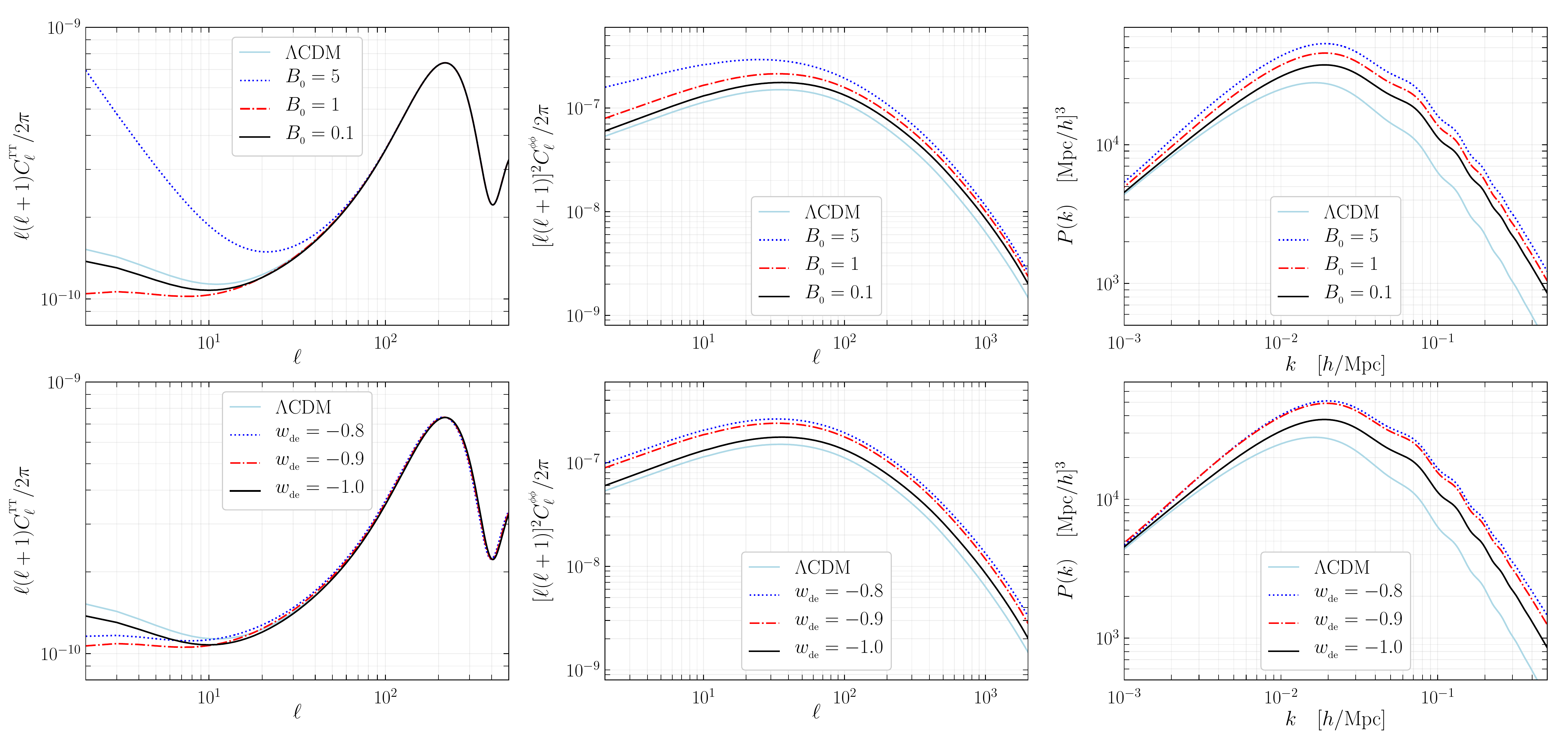}
  \caption[adjust]{Effects of $f_{\mathcal{R}}$ gravity on the CMB angular temperature power spectrum (left), lensing 
  power spectrum (middle) and the linear matter power spectrum (right) for different designer $f(\mathcal{R})$ models 
  against the $\Lambda$CDM predictions. Unless otherwise written, we chose $\qsubrm{w}{de}=-1$ and $\qsubrm{B}{0}=0.1$ 
  as well as the same cosmology as in Fig.~\ref{fig:deltade}.}
  \label{fig:obs}
 \end{center}
\end{figure*}

The CMB angular anisotropy power spectrum is a snapshot of the acoustic waves in the photon-baryon fluid at decoupling, 
distorted by the Integrated Sachs-Wolfe effect (ISW) and the lensing due to the subsequent gravitational collapse of 
the matter. How and when can dark energy perturbations in $f(\mathcal{R})$ gravity affect the CMB anisotropy? 
Since in viable $f(\mathcal{R})$ gravity models, dark energy perturbations are subdominant at early time (see point 
\ref{point2} on page~\pageref{point2}), they can not have any impact on the physical phenomena at play at the epoch of 
decoupling. However, they alter the growth of structure from the end of the matter dominated era (see point 
\ref{point3} on page~\pageref{point3}). Therefore, they may have an impact on the \textit{late} ISW effect (see, e.g. 
\citep{Lesgourgues2013}) and lensing of the CMB anisotropy (see, e.g. \citep{Lewis2006}). The late ISW effect is 
contributing to the CMB temperature anisotropy on large angular scales ($\ell\lesssim 20$) and the lensing power 
spectrum of the CMB probes structure formation on a wider range of scales ($\ell\lesssim 1000$). So we expect the CMB 
angular anisotropy power spectrum to be affected by dark energy perturbations only at low multipoles, i.e., where the 
cosmic variance limits the constraining power of the CMB data. Hence, the lensing power spectrum shall be a more 
compelling probe of dark energy perturbation than the CMB temperature anisotropy angular power spectrum.

\begin{table*}[!ht]
 \begin{center}
  \renewcommand*{\arraystretch}{1.2}
  \caption{\label{tab:con1} Posterior mean (68\% C.L.) for $\log{\qsubrm{B}{0}}$, $\qsubrm{\sigma}{8}$ and 
  $\qsubrm{w}{de}$ for designer $f(\mathcal{R})$ models that mimic a $\Lambda$CDM and a $w$CDM expansion. The ellipses 
  indicate the absence of 68\% C.L. constraints, in this case only the 95\% C.L. upper limits are relevant (see Table 
  \ref{tab:con2}).}
  \begin{ruledtabular}
   \begin{tabular}{ccccc}
    & CMB+BAO                & CMB+BAO+Lensing & CMB+BAO  & CMB+BAO+Lensing\\
    & ($\Lambda$CDM)         & ($\Lambda$CDM)  & ($w$CDM) & ($w$CDM)\\
    \hline
    $\log{\qsubrm{B}{0}}$    & $-2.01^{+1.26}_{-0.19}$ & $\cdots$ & $\cdots$ & $\cdots$ \\
    
    $\qsubrm{\sigma}{8}$     & $1.04^{+0.10}_{-0.03}$ & $\cdots$ & 
                               $1.13^{+0.05}_{-0.03}$ & $0.98^{+0.05}_{-0.03}$\\
    
    $(1+\qsubrm{w}{de})\times10^3$         & $0$  & $0$  & $8.10^{+1.50}_{-8.10}$ & $0.64^{+0.08}_{-0.64}$ \\
   \end{tabular}
  \end{ruledtabular}
 \end{center}
\end{table*}

\vspace{0.5cm}

\begin{table*}[!ht]
 \begin{center}
  \renewcommand*{\arraystretch}{1.2}
  \caption{\label{tab:con2} Posterior upper limits (95\% C.L.) for $\log{\qsubrm{B}{0}}$, $\qsubrm{\sigma}{8}$ and 
  $\qsubrm{w}{de}$ for designer $f(\mathcal{R})$ models that mimic a $\Lambda$CDM and a $w$CDM expansion.}
  \begin{ruledtabular}
   \begin{tabular}{ccccc}
    & CMB+BAO                & CMB+BAO+Lensing & CMB+BAO  & CMB+BAO+Lensing\\
    & ($\Lambda$CDM)         & ($\Lambda$CDM)  & ($w$CDM) & ($w$CDM)\\
    \hline
    $\log{\qsubrm{B}{0}}$    & $<-0.78$   & $<-2.2$ & $<-1.26$ & $<-2.35$ \\
    
    $\qsubrm{\sigma}{8}$     & $<1.13$ & $<0.99$ & $<1.18$ & $<1.04$\\
    
    $(1+\qsubrm{w}{de})\times10^3$         & $0$   & $0$   & $<20$ & $<2.1$ \\
   \end{tabular}
  \end{ruledtabular}
 \end{center}
\end{table*}

In the left panels of Fig.~\ref{fig:obs} we show the CMB temperature angular anisotropy power spectrum computed for 
several designer models with different $\qsubrm{w}{de}$ and $\qsubrm{B}{0}$, against the $\Lambda$CDM prediction. 
We see that significant differences appear when $\qsubrm{B}{0}\gtrsim 1$ and that the late ISW effect can be strongly 
enhanced for larger values of $\qsubrm{B}{0}$. Moreover, at fixed $\qsubrm{B}{0}$ the late ISW contribution is more 
significant for less negative $\qsubrm{w}{de}$, as can be understood with the results of Sec.~\ref{sect:results} 
(see point \ref{point3} on page~\pageref{point3}). In the middle panels we show the CMB lensing power spectrum computed 
in the same settings. Its amplitude is larger for larger $\qsubrm{B}{0}$ and less negative $\qsubrm{w}{de}$, again in 
agreement with the analysis of Sec.~\ref{sect:results}. Similar conclusions apply to the linear matter power 
spectrum presented in the right panels of Fig.~\ref{fig:obs}. In particular, for scales which are still in the GR 
regime today ($k\approx 10^{-3} h\,\mathrm{Mpc}^{-1}$), the amplitude of the matter power spectrum is close to the 
$\Lambda$CDM prediction, while for scales that entered the ST regime during the matter dominated era 
($k\gtrsim 10^{-2} h\,\mathrm{Mpc}^{-1}$), its amplitude is enhanced.

\begin{table}[!th]
 \begin{center}
  \renewcommand*{\arraystretch}{1.2}
  \caption{\label{tab:con3} Posterior mean (68\% C.L.) for $\qsubrm{\sigma}{8}$ and $\qsubrm{w}{de}$ for a $w$CDM 
  model.}
  \begin{ruledtabular}
   \begin{tabular}{ccc}
    & CMB+BAO & CMB+BAO+Lensing \\
    \hline
    $\qsubrm{\sigma}{8}$     & $0.85^{+0.02}_{-0.02}$ & $0.83^{+0.02}_{-0.02}$\\
    
    $(1+\qsubrm{w}{de})\times10^2$         & $-7.35^{+7.76}_{-5.9}$ & $-4.7^{+6.5}_{-6.1}$ \\
   \end{tabular}
  \end{ruledtabular}
 \end{center}
\end{table}
   
\vspace{0.5cm}

\begin{table}[!th]
 \begin{center}
  \renewcommand*{\arraystretch}{1.2}
  \caption{\label{tab:con4} Posterior upper limits (95\% C.L.) for $\qsubrm{\sigma}{8}$ and $\qsubrm{w}{de}$ for a 
  $w$CDM model.}
  \begin{ruledtabular}
   \begin{tabular}{ccc}
    & CMB+BAO & CMB+BAO+Lensing \\
    \hline
    $\qsubrm{\sigma}{8}$     & $<0.89$ & $<0.87$\\
    
    $(1+\qsubrm{w}{de})\times10^2$         & $<5.9$ & $<8$ \\
   \end{tabular}
  \end{ruledtabular}
 \end{center}
\end{table}

For observational constraints, we consider the following combinations of data sets: CMB+BAO and CMB, BAO+Lensing. 
For CMB and Lensing we refer to the Planck 2015 public likelihoods for low-$\ell$ and high-$\ell$ temperature as 
well as polarisation and lensing data \citep{Planck2016_XIII}. 
For BAO we refer to the distance measurements provided by the WiggleZ Dark Energy Survey \citep{Kazin2014} and SDSS 
\citep{Ross2015}. We use \verb|Montepython| \citep{Audren2013} for the Monte Carlo Markov chain sampling of the 
parameter space. We varied the six base cosmological parameters as well as all the Planck nuisance parameters. For 
those, we used the same priors as the Planck Collaboration \citep{Planck2016_XIII}. In addition we varied the 
background dark energy equation of state $\qsubrm{w}{de}$ and $\log{\qsubrm{B}{0}}$ that characterise the designer 
$f(\mathcal{R})$ models. For $\qsubrm{w}{de}$ we used a uniform prior between $-1$ and $0$. For 
$\log{\qsubrm{B}{0}}$ we used a uniform prior between $-6$ and $1$. In Tables~\ref{tab:con1} and ~\ref{tab:con2}, we 
show the 68\% C.L. and 95\% C.L. constraints from our analyses.

For designer models with $\qsubrm{w}{de}=-1$, $\qsubrm{B}{0}$ and $\qsubrm{\sigma}{8}$ are determined at 68\% C.L. for 
CMB+BAO. We get $\qsubrm{B}{0}\approx0.01$ and $\qsubrm{\sigma}{8}\simeq 1.0\pm0.1$. If we add the information relative 
to clustering at late time, via the CMB lensing data, $\qsubrm{B}{0}$ and $\qsubrm{\sigma}{8}$ are not determined, but  
constrained to $\qsubrm{B}{0}\lesssim 0.006$ and $\qsubrm{\sigma}{8} < 1.0$ (95\% C.L.).

For designer models with $\qsubrm{w}{de}\neq-1$, $\qsubrm{B}{0}$ is not determined any more by CMB+BAO. Moreover, due 
to the correlation between $\qsubrm{w}{de}$ and the amplitude of clustering (see bottom panel of 
Fig.~\ref{fig:deltade}), $\qsubrm{\sigma}{8}$ takes substantially larger values than with the $\qsubrm{w}{de}=-1$ 
models. 
When we add CMB lensing data, the posterior mean value of $\qsubrm{\sigma}{8}$ is brought down by fifteen percent and 
more importantly the 68\% C.L. region for the dark energy background equation of state is reduced by a factor of ten. 
We get $(1+\qsubrm{w}{de})<0.0006$, in other words the expansion history has to be very close to $\Lambda$CDM.

\section{Comparison with \texorpdfstring{$w$CDM}{wCDM} models}\label{sect:wCDM}
To quantify the relative importance of perturbations in (designer) $f(\mathcal{R})$ models, we can compare their 
observational constraints with a $w$CDM model where the background equation of state $w$ is free to vary (but 
constant in time) and we keep the sound speed (defined in the frame comoving with the fluid) 
$c_{\rm s}^2 = \delta p/\delta\rho = 1$ fixed. To study the perturbations of such a model, we use the \verb|CLASS| 
implementation of the parameterized post-Friedmaniann (PPF) framework as described in \cite{Fang2008}. When 
$\qsubrm{w}{de}\geq -1$, this framework recovers the behaviour of canonical minimally coupled scalar field models and 
it is accurate also when $\qsubrm{w}{de}\approx -1$. A welcome aspect of the PPF formalism is that it allows to study 
the evolution of perturbations in the phantom regime ($\qsubrm{w}{de}<-1$), which is usually preferred by Supernovae 
data \citep{Tonry2003,Riess2004}. In addition, the crossing of the ``phantom barrier" ($\qsubrm{w}{de}=-1$) is allowed, 
covering therefore also the more general case of non-canonical minimally coupled models, such as $k$-essence. The PPF 
formalism allows also sound speeds $c_{\rm s}^2\neq 1$, as in $k$-essence models, but here we limit ourselves to the 
standard case of luminal sound speed, as this is also the value in $f(\mathcal{R})$ models.

We note, in principle, that in $w$CDM models $\qsubrm{w}{de}$ can take values smaller than -1, this is the so-called 
phantom regime, while in the designer $f(\mathcal{R})$ models we consider in this work the phantom crossing is not 
allowed due to instabilities, see Sec.~\ref{sect:results}.

Moreover we saw that in $f(\mathcal{R})$ gravity small variations of $\qsubrm{w}{de}$ lead to large variations in 
$\qsubrm{\sigma}{8}$ (see bottom panel of figure~\ref{fig:deltade}), while in $w$CDM models small variations of 
$\qsubrm{w}{de}$ lead to small variations in $\qsubrm{\sigma}{8}$: in the range of $\qsubrm{w}{de}$ presented in the 
bottom panel of figure~\ref{fig:deltade}, for the same cosmological parameters, $\qsubrm{\sigma}{8}$ would vary by less 
than 1\%.

Using the same data sets described before, in Tables~\ref{tab:con3} and \ref{tab:con4} we show the 68\% and 95\% C.L. 
constraints on $\qsubrm{\sigma}{8}$ and $\qsubrm{w}{de}$ for the $w$CDM fluid, respectively. For $\qsubrm{w}{de}$, we 
use a uniform prior between $-2$ and $0$.

Our results agree with \cite{Planck2016_XIII}. In particular, the preferred value for $\qsubrm{w}{de}$ is in the 
phantom regime. It means that these data sets favour a higher value of $\qsubrm{\sigma}{8}$ with respect to the 
$\Lambda$CDM cosmology, as was the case for the $f(\mathcal{R})$ models.

Our last remark is that since $\qsubrm{\sigma}{8}$ depends weakly on $\qsubrm{w}{de}$ in $w$CDM compared to 
$f(\mathcal{R})$, the constraints on $\qsubrm{w}{de}$ in $w$CDM are weaker than in $f(\mathcal{R})$ by one order of 
magnitude, see tables~\ref{tab:con3} and \ref{tab:con4}.

\section{Discussion and conclusion}\label{sect:conclusion}
Intense observational and theoretical efforts are being deployed to unveil the nature of the cosmic acceleration of the 
universe. Going beyond the cosmological constant $\Lambda$, two main hypotheses can be explored: dark energy and 
modified gravity. Many models belonging to these two broad groups can be described in terms of the Horndeski 
Lagrangian. In this work we concentrated on a well studied sub-class of Horndeski theories, the so-called 
$f(\mathcal{R})$ gravity models. Such modifications to GR may affect both the background expansion history and the 
evolution of cosmological perturbations. 
In this paper we considered the designer $f(\mathcal{R})$ gravity models for which the $f(\mathcal{R})$ function is 
tuned to reproduce the $w$CDM expansion history.

We used the EoS approach to study analytically the dynamics of linear cosmological perturbations in this context, and 
we implemented it numerically in our \verb|CLASS_EOS_FR| code. 
To prove the reliability of our numerical implementation, we compared our results with several other $f(\mathcal{R})$ 
codes publicly available such as \verb|MGCAMB| \citep{Zhao2009a,Hojjati2011}, \verb|FRCAMB| \citep{He2012}, 
\verb|EFTCAMB| \citep{Hu2014a,Raveri2014,Hu2014b} and found agreement at the sub-percent level for all of them 
\citep{Bellini2018}, except for \verb|MGCAMB| which disagreed by more than five percent relative error with the other 
codes for the computation of the matter power spectrum for $k> 1~h\,\mathrm{Mpc}^{-1}$.

Unlike for the simple $w$CDM dark energy model, we found that for designer $f(\mathcal{R})$ gravity models a less 
negative $\qsubrm{w}{de}$ leads to a larger $\qsubrm{\sigma}{8}$ (see point \ref{point3} on page~\pageref{point3}). To 
arrive at this conclusion we derived an analytical formula for the growth index $\gamma$ (see Eq. \eqref{eqn:gammaST}).

Using CMB lensing data we found that designer $f(\mathcal{R})$ models with $(1+\qsubrm{w}{de})>0.002$ and 
$\qsubrm{B}{0}>0.006$ are disfavoured at 95\% C.L. Note that similar constraints were obtained for the designer 
$f(\mathcal{R})$ models also by \cite{Raveri2014}, using cosmological data as we did here. 
The authors of \cite{Hu2016} performed a similar analysis on the Hu-Sawicki $f(\mathcal{R})$ models and found, as we 
did, a higher value of $\qsubrm{\sigma}{8}$ with respect to the $\Lambda$CDM value\footnote{Using CFHTLenS data 
\citep{Heymans2013}, the normalisation of the matter power spectrum is significantly closer to the $\Lambda$CDM value, 
implying a lower value of $f_{\mathcal{R},0}$ ($\qsubrm{B}{0}$ in our notation) and, using their Eq.~10 (see also their 
figure~2), $\qsubrm{w}{de}\approx -1$.}. Moreover, for the screening mechanism to happen on solar system scales the 
authors of \citep{Brax2008,Faulkner2007} found $|1+\qsubrm{w}{de}|\lesssim 10^{-4}$ for generic $f(\mathcal{R})$ 
models.\\

The results we obtained are consistent with these previous analyses and hint for the fact that generic $f(\mathcal{R})$ 
models with $\qsubrm{w}{de}\neq-1$ can be ruled out based on current cosmological data, complementary to solar system 
tests.

\section{Acknowledgements}
We thank Ruth Durrer, Marco Raveri, Alessandra Silvestri, Filippo Vernizzi, Bin Hu, Jens Chluba and Lucas Lombriser for 
discussions. We thank Julien Lesgourgues, Thomas Tram and Thejs Brinckmann for their help with CLASS and Montepython. 
FP acknowledges financial support from the STFC Grant R120562. BB acknowledges financial support from the ERC 
Consolidator Grant 725456. Part of the analysis presented here is based on observations obtained with Planck 
(http://www.esa.int/Planck), an ESA science mission with instruments and contributions directly funded by ESA Member 
States, NASA, and Canada. We also thank the referee whose comments helped us to improve the scientific content of 
this work.

\appendix*

\section{Comparison between the EoS and EFT approaches for dark energy perturbations}\label{sect:EoS_EFT}
In this section we compare the expressions for the entropy perturbations and the perturbed anisotropic stress of 
\cite{Battye2016a} with the corresponding expressions from \cite{Gleyzes2014} in the conformal Newtonian gauge. In the 
following we will denote with the superscript ``BBP" variables in \cite{Battye2016a} and with``GLV" variables in 
\cite{Gleyzes2014}. In addition we use the subscript `m' for all matter species and 
\begin{equation}
\qsubrm{\zeta}{i}=\tfrac{\qsubrm{g}{K}\qsubrm{\epsilon}{H}-\qsubrm{\bar{\epsilon}}{H}}
{3\qsubrm{g}{K}\qsubrm{\epsilon}{H}}-\tfrac{dP_i}{d\rho_i}.
\end{equation}
In $f(\mathcal{R})$ gravity the equations of states for scalar perturbations, in both formalisms are 
\cite{Battye2016a,Gleyzes2014}
\begin{widetext}
 \bse
  \label{eq:eos_fr}
  \bea
   \qsubrm{\Pi}{de}^{{\scriptscriptstyle{\textrm{BBP}}}} & = & 
   \frac{\mathrm{K}^2}{3\qsubrm{g}{K}\qsubrm{\epsilon}{H}}\left\{\qsubrm{\Delta}{de} - 
   \frac{f_{\mathcal{R}}^\prime}{2(1+f_\mathcal{R})}\qsubrm{\Theta}{de}
   +\frac{\qsubrm{\Omega}{m}}{\qsubrm{\Omega}{de}}\frac{f_{\mathcal{R}}}{1+f_\mathcal{R}}\qsubrm{\Delta}{m} - 
   \frac{\qsubrm{\Omega}{m}}{\qsubrm{\Omega}{de}}\frac{f_{\mathcal{R}}^\prime}{2(1+f_\mathcal{R})}
   \qsubrm{\Theta}{m}\right\} - 
   \frac{f_{\mathcal{R}}}{1+f_{\mathcal{R}}}\frac{\qsubrm{\Omega}{m}}{\qsubrm{\Omega}{de}}
  \qsubrm{\Pi}{m},\quad\quad \label{eq:eosPi_S}\\
   \qsubrm{\Gamma}{de}^{{\scriptscriptstyle{\textrm{BBP}}}} & = & 
   \left\{\qsubrm{\zeta}{de}-\frac{\qsubrm{\bar{\epsilon}}{H}}{3\qsubrm{g}{K}\qsubrm{\epsilon}{H}}
   \frac{2(1+f_\mathcal{R})-f_\mathcal{R}^\prime}
   {f_\mathcal{R}^\prime}\right\}\qsubrm{\Delta}{de} - 
   \qsubrm{\zeta}{de}\qsubrm{\Theta}{de}
   +\frac{\qsubrm{\Omega}{m}}{\qsubrm{\Omega}{de}}
   \left\{\qsubrm{\zeta}{m}-\frac{\qsubrm{\bar{\epsilon}}{H}}{3\qsubrm{g}{K}\qsubrm{\epsilon}{H}}
   \frac{2f_\mathcal{R}-f_\mathcal{R}^\prime}{f_\mathcal{R}^\prime}\right\}
   \qsubrm{\Delta}{m}
   -\frac{\qsubrm{\Omega}{m}}{\qsubrm{\Omega}{de}}\qsubrm{\zeta}{m}\qsubrm{\Theta}{m} - 
   \frac{\qsubrm{\Omega}{m}}{\qsubrm{\Omega}{de}}\qsubrm{\Gamma}{m},\\
   \label{eq:eosGamma}
  \label{eqn:eos_fr_GLV}
   \qsubrm{P}{de}^{{\scriptscriptstyle{\textrm{GLV}}}}\qsubrm{\Gamma}{de}^{{\scriptscriptstyle{\textrm{GLV}}}} & = & 
                 \frac{\qsubrm{\gamma}{1}\qsubrm{\gamma}{2} + \qsubrm{\gamma}{3} \qsubrm{\alpha}{B}^2\rm{K}^2}
                 {\qsubrm{\gamma}{1} + \qsubrm{\alpha}{B}^2 \rm{K}^2}
                 (\delta\qsubrm{\rho}{de}-3H\qsubrm{q}{de})
                 +\frac{\qsubrm{\gamma}{1}\qsubrm{\gamma}{4}+\qsubrm{\alpha}{B}^2\rm{K}^2}
                 {\qsubrm{\gamma}{1}+\qsubrm{\alpha}{B}^2\rm{K}^2}H
                 (\qsubrm{q}{de}+\qsubrm{q}{m})
                 +\frac{1}{3}(\delta\qsubrm{\rho}{m} - 3H\qsubrm{q}{m})
                 -\frac{\qsubrm{dP}{de}}{\qsubrm{d\rho}{de}}\qsubrm{\delta\rho}{de}
                 -\qsubrm{\delta p}{m}, \label{eqn:GammaGLV}\\
   \qsubrm{P}{de}^{{\scriptscriptstyle{\textrm{GLV}}}}\qsubrm{\Pi}{de}^{{\scriptscriptstyle{\textrm{GLV}}}} & = & 
                 \frac{\qsubrm{\gamma}{8} \qsubrm{\alpha}{B}^2 \rm{K}^2}
                 {2(\qsubrm{\gamma}{1} + \qsubrm{\alpha}{B}^2 \rm{K}^2)}
                 (\qsubrm{\delta\rho}{de}-3H\qsubrm{q}{de})
                 -\frac{\qsubrm{\gamma}{9} \rm{K}^2}{2(\qsubrm{\gamma}{1} + \qsubrm{\alpha}{B}^2\rm{K}^2)}H
                 (\qsubrm{q}{de}+\qsubrm{q}{m}),\label{eqn:PiGLV}
  \eea
 \ese
\end{widetext}
where the functions $\gamma_i$ are given by $\qsubrm{\gamma}{1} = 3\qsubrm{\alpha}{B}^2\qsubrm{\epsilon}{H}$, 
$\qsubrm{\gamma}{3} = \tfrac{1}{3}$, 
$\qsubrm{\gamma}{2} = \tfrac{1}{3}-\frac{\qsubrm{\bar{\epsilon}}{H}}{3\qsubrm{\epsilon}{H}\qsubrm{\alpha}{B}}$, 
$\qsubrm{\gamma}{4}=1-\tfrac{\qsubrm{\bar{\epsilon}}{H}}{\qsubrm{\epsilon}{H}}$, $\qsubrm{\gamma}{8} = -2$, 
$\qsubrm{\gamma}{9} = -6\qsubrm{\alpha}{B}^3$. We further define  
$\qsubrm{\alpha}{B} = \tfrac{f^{\prime}_{\mathcal{R}}}{2(1+f_{\mathcal{R}})}$. Note that with respect to 
\cite{Gleyzes2014}, we defined 
$\qsubrm{P}{de}\qsubrm{\Pi}{de}^{{\scriptscriptstyle{\textrm{GLV}}}}=
-\tfrac{k^2}{a^2}\qsubrm{\sigma}{de}^{{\scriptscriptstyle{\textrm{GLV}}}}$.\\
\\
\\
\\
Unlike \cite{Battye2016}, the authors of \cite{Gleyzes2014} use a non-standard continuity equation for the effective 
dark energy fluid which implies
\begin{align}
 \qsubrm{\rho}{de}^{{\scriptscriptstyle{\textrm{GLV}}}} & = 
 \qsubrm{\rho}{de}^{{\scriptscriptstyle{\textrm{BBP}}}}+3\qsubrm{M}{pl}^2H^2f_{\mathcal{R}}\;,\nonumber\\
 \qsubrm{P}{de}^{{\scriptscriptstyle{\textrm{GLV}}}} & = 
 \qsubrm{P}{de}^{{\scriptscriptstyle{\textrm{BBP}}}}-
 \qsubrm{M}{pl}^2H^2\left(3-2\qsubrm{\epsilon}{H}\right)f_{\mathcal{R}}\;,\nonumber
\end{align}
for the background and
\begin{align}
 \qsubrm{\delta\rho}{de}^{{\scriptscriptstyle{\textrm{GLV}}}} & = 
                                   (1+f_{\mathcal{R}})
                                   \qsubrm{\delta\rho}{de}^{{\scriptscriptstyle{\textrm{BBP}}}}+
                                   f_{\mathcal{R}}
                                   \qsubrm{\delta\rho}{m}^{{\scriptscriptstyle{\textrm{BBP}}}}\;,\nonumber\\
 \qsubrm{\delta P}{de}^{{\scriptscriptstyle{\textrm{GLV}}}} & = 
                                 (1+f_{\mathcal{R}})\qsubrm{\delta P}{de}^{{\scriptscriptstyle{\textrm{BBP}}}}+
                                 f_{\mathcal{R}}\qsubrm{\delta P}{m}^{{\scriptscriptstyle{\textrm{BBP}}}}\;,\nonumber\\
 \qsubrm{q}{m}^{{\scriptscriptstyle{\textrm{GLV}}}} + 
 \qsubrm{q}{de}^{{\scriptscriptstyle{\textrm{GLV}}}} & = 
                                    -\frac{1+f_{\mathcal{R}}}{3H}\left\{
                                    \qsubrm{\rho}{de}^{{\scriptscriptstyle{\textrm{BBP}}}}
                                    \qsubrm{\Theta}{de}^{{\scriptscriptstyle{\textrm{BBP}}}}+
                                    \qsubrm{\rho}{m}^{{\scriptscriptstyle{\textrm{BBP}}}}
                                    \qsubrm{\Theta}{m}^{{\scriptscriptstyle{\textrm{BBP}}}}\right\}\;,
                                    \nonumber\\
  \qsubrm{q}{de}^{{\scriptscriptstyle{\textrm{GLV}}}} & = 
                           -\frac{1}{3H}\left\{
                           (1+f_{\mathcal{R}})\qsubrm{\rho}{de}^{{\scriptscriptstyle{\textrm{BBP}}}}
                           \qsubrm{\Theta}{de}^{{\scriptscriptstyle{\textrm{BBP}}}}+
                           f_{\mathcal{R}}\qsubrm{\rho}{m}^{{\scriptscriptstyle{\textrm{BBP}}}}
                           \qsubrm{\Theta}{m}^{{\scriptscriptstyle{\textrm{BBP}}}}\right\}\;,
                           \nonumber\\
 \qsubrm{P}{de}^{{\scriptscriptstyle{\textrm{GLV}}}}
 \qsubrm{\Pi}{de}^{{\scriptscriptstyle{\textrm{GLV}}}} & = 
                                    \left[(1+f_{\mathcal{R}})
                                    \qsubrm{P}{de}^{{\scriptscriptstyle{\textrm{BBP}}}}
                                    \qsubrm{\Pi}{de}^{{\scriptscriptstyle{\textrm{BBP}}}}+
                                    f_{\mathcal{R}}
                                    \qsubrm{P}{m}^{{\scriptscriptstyle{\textrm{BBP}}}}
                                    \qsubrm{\Pi}{m}^{{\scriptscriptstyle{\textrm{BBP}}}}\right]\;. \nonumber
\end{align}
for the perturbed fluid variables. From this, we conclude that both formalisms are equivalent. 

\bibliographystyle{apsrev4-1}
\bibliography{EoSForDE.bbl}

\end{document}